\documentclass[prl,aps,amssymb,amsmath]{article}
\usepackage{amssymb,amsmath,tikz}
\usepackage{amsthm}
\usepackage{mathrsfs}
\usepackage{pifont}

\showoutput
\renewcommand{\qedsymbol}{$\blacksquare$}

\newcommand{\ud}{\mathrm{d}}

\theoremstyle{plain}

\newtheorem*{twr*}{THEOREM}
\newtheorem*{lem*}{LEMMA}

\newtheorem*{rem*}{REMARK}

\newtheorem*{notn*}{NOTATION}
\newtheorem*{wiener-ito*}{WIENER-IT\^O-SEGAL DECOMPOSITION}

\begin{document}
\title{ {\bf A new theorem on the representation structure of the $SL(2, \mathbb{C})$ 
group acting in the Hilbert space of the quantum Coulomb field}}
\author{Jaros{\l}aw Wawrzycki \footnote{Electronic address: jaroslaw.wawrzycki@wp.pl or jaroslaw.wawrzycki@ifj.edu.pl}
\\Institute of Nuclear Physics of PAS, ul. Radzikowskiego 152, 
\\31-342 Krak\'ow, Poland}
\maketitle

\vspace{1cm}

\begin{abstract}
Using the results obtained by Staruszkiewicz in 
\emph{Acta Phys. Pol. {\bf B 23}, 591 (1992)} and in \emph{Acta Phys. Pol. {\bf B 23}, 927 (1992)}
 we show  that the representations acting in the eigenspaces of the total charge operator corresponding to the 
eigenvalues $n_1, n_2$ whose absolute values are less than or equal $\sqrt{\pi/e^2}$ are inequivalent
if $|n_1| \neq |n_2|$
and contain the supplementary series component acting as a discrete component. On the other hand the representations
acting in the eigenspaces corresponding to eigenvalues whose absolute values are greater than $\sqrt{\pi/e^2}$
are all unitarily equivalent and do not contain any supplementary series component.
\end{abstract}

\section{Introduction}

In this paper we prove a new theorem within the Quantum Theory of the Coulomb Field,
\cite{Staruszkiewicz1987}, \cite{Staruszkiewicz}. This paper can be regarded as an immediate continuation of the series
of Staruszkiewicz's papers \cite{Staruszkiewicz1992ERRATUM}, \cite{Staruszkiewicz1992}, \cite{Staruszkiewicz2004},
on the structure of the unitary representation of  
$SL(2, \mathbb{C})$ acting in the Hilbert space of the quantum Coulomb field and the quantum phase field
$S(x)$ of his theory, and its connection to the fine structure constant. We use the notation of these papers.
Basing on the results of these papers we give here a proof of the following 

\vspace*{0.5cm}

THEOREM.
Let $U|_{{}_{\mathcal{H}_{{}_{m}}}}$ be the restriction of the unitary representation $U$ of $SL(2, \mathbb{C})$
in the Hilbert space of the quantum phase field $S$ to the invariant eigenspace ${\mathcal{H}_{{}_{m}}}$
of the total charge operator $Q$ corresponding to the eigenvalue $me$ for some integer $m$. Then 
for all $m$ such that
\[
|m| > \textrm{Integer part} \Big( \sqrt{\frac{\pi}{e^2}} \Big)
\]
the representations $U|_{{}_{\mathcal{H}_{{}_{m}}}}$ are unitarily equivalent:
\[
U|_{{}_{\mathcal{H}_{{}_{m}}}} \cong_{{}_{U}}
U|_{{}_{\mathcal{H}_{{}_{m'}}}}
\]
whenever 
\[
|m| > \textrm{Integer part} \Big( \sqrt{\frac{\pi}{e^2}} \Big), \,\,\,
|m'| > \textrm{Integer part} \Big( \sqrt{\frac{\pi}{e^2}} \Big).
\]
 
On the other hand if the two 
integers $m,m'$ have different absolute values $|m| \neq |m'|$ and are such that 
\[
|m| <  \sqrt{\frac{\pi}{e^2}}, \,\,\,
|m'| < \sqrt{\frac{\pi}{e^2}},
\]
then the representations $U|_{{}_{\mathcal{H}_{{}_{m}}}}$ and $U|_{{}_{\mathcal{H}_{{}_{m'}}}}$
are inequivalent. Each representation $U|_{{}_{\mathcal{H}_{{}_{m}}}}$ contains a unique discrete 
supplementary component if
\[
|m| <  \sqrt{\frac{\pi}{e^2}}, \,\,\,
\] 
and the supplementary components contained in $U|_{{}_{\mathcal{H}_{{}_{m}}}}$ with different values of $|m|$ fulfilling the last inequality are inequivalent. If
\[
|m| > \textrm{Integer part} \Big( \sqrt{\frac{\pi}{e^2}} \Big)
\]
then the representation $U|_{{}_{\mathcal{H}_{{}_{m}}}}$ does not contain in its decomposition any 
supplementary components. \qed

\vspace*{0.5cm}

This remarkable result can be compared to the well known and curious coincidence concerning
self-adjointness of the Hamiltonian of the bounded system composed by a heavy source (say nucleus) of the 
classical Coulomb 
field and a relativistic charged particle in this field. Namely it is a well known phenomenon in 
relativistic wave mechanics that whenever 
the charge of the nuclei is of the order of magnitude comparable to the inverse of the fine structure
constant or greater, then the Hamiltonian loses the self-adjointness property (which sometimes
is interpretad as an indication that the system, when passing to the quantum field theory level, becomes unstable).
 On the other hand (and this is a coincidence which no one understands) the nuclei of real atoms are unstable 
whenever the charge of the nuclei reaches the 
value of the same order (inverse of the fine structure constant). The mentioned breakdown of self-adjointness
 cannot explain of course this phenomenon because there are mostly the strong (and not electromagnetic) forces which govern the stability of nuclei. To this coincidence we add another
coming from the quantum theory of infrared photons of the quantized Coulomb field. Although we should emphasize that the 
mentioned three phenomena come from three different regimes and so far we are not able to unswer the question if these coincidences are merely accidental or not.

\section{Proof of the theorem}

Let us concentrate our attention on the specific state $|u\rangle$ in the eigenspace $\mathcal{H}_{{}_{m=1}}$ corresponding to the eigenvalue $e$ of the charge operator $Q$. 
For any time like unit vector $u$ we can form the following unit vector (compare \cite{Staruszkiewicz1992ERRATUM}
or \cite{Staruszkiewicz2004})
\begin{equation}\label{kernel-AS}
|u\rangle = e^{-iS(u)} |0\rangle
\end{equation} 
in the Hilbert space of the quantum field $S$. It has the following properties
\begin{enumerate}
\item[1)]
$|u\rangle$ is an eigenstate of the total charge $Q$: $Q|u\rangle = e |u\rangle$.
\item[2)]

$|u\rangle$ is spherically symmetric in the rest frame of $u$: 
$\epsilon^{\alpha \beta \mu \nu}u_{\beta} M_{\mu \nu} |u\rangle  = 0$, where $M_{\mu \nu}$
are the generators of the $SL(2, \mathbb{C})$ group.
\item[3)]
$|u\rangle$ does not contain the (infrared) transversal photons: $N(u) |u\rangle = 0$, where
$N(u)$ is the operator of the number of transversal photons in the rest frame of $u$.
If $u$ is the four-velocity of the reference frame in which the partial waves $f_{lm}^{(+)}$ 
are computed, then in this reference frame
\[
N(u) = (4\pi e^2)^{-1} \sum \limits_{l=1}^{\infty} \sum \limits_{m = -l}^{l} c_{lm}^{+}c_{lm},
\]
and (up to an irrelevant phase factor)
\[
|u\rangle = e^{-iS_0}|0\rangle.
\]
\end{enumerate}

These three conditions determine the state vector $|u\rangle$ up to a phase factor.

Now let us consider the subspace 
$\mathcal{H}_{{}_{|u\rangle}} \subset \mathcal{H}_{{}_{m=1}}$ as spanned by the vectors of the form 
$U_{{}_{\alpha}}|u\rangle$, $\alpha \in SL(2,\mathbb{C})$.

Note that the above conditions 1) and 2) determine $|u\rangle$ as the ``maximal''
vector in $\mathcal{H}_{{}_{|u\rangle}}$ which preserves the conditions 1), 2), i.e. any state vector in the Hilbert 
subspace $\mathcal{H}_{{}_{|u\rangle}}$ of the quantum phase field 
$S$ which preserves 1) and 2) and which is orthogonal to $|u\rangle$ is equal zero. 

First: in the paper \cite{Staruszkiewicz} it was computed that the inner product
\[
\langle u| v \rangle = \exp\Big\{ - \frac{e^2}{\pi}(\lambda \textrm{coth} \lambda - 1) \Big\},
\]
where $u\cdot v = g_{\mu \nu}u^\mu v^\mu = \textrm{cosh} \lambda$, so that $\lambda$ is the hyperbolic angle between 
$u$ and $v$; compare also \cite{Staruszkiewicz2002}.  

Second: it was proved in \cite{Staruszkiewicz1992ERRATUM} (compare also \cite{Staruszkiewicz2004}, 
\cite{Staruszkiewicz2009}) that the state $|u\rangle$, lying in the subspace $Q = e \boldsymbol{1}$ of the

Hilbert space of the field $S$, when decomposed 
into components corresponding to 
the decomposition of $U$ into irreducible sub-representations contains
\begin{enumerate}
\item[-] only the principal series if $\frac{e^2}{\pi} >1$,
\item[-] the principal series and a discrete component from the supplementary series with
\[
-\frac{1}{2}M_{\mu \nu}M^{\mu \nu} = z(2-z) \boldsymbol{1}, \,\,\, z = \frac{e^2}{\pi} ,
\]
if $0 < \frac{e^2}{\pi} < 1$,
\end{enumerate}   
in the units in which $\hslash = c = 1$. In other units one should read $\frac{e^2}{\pi \hslash c}$
for $\frac{e^2}{\pi}$.

In particular from the result of \cite{Staruszkiewicz1992ERRATUM} it follows that for the restriction $U|_{{}_{\mathcal{H}_{{}_{|u\rangle}}}}$ of the representation
$U$ of $SL(2, \mathbb{C})$ acting in the Hilbert space of the quantum ``phase'' field $S$ to the invariant subspace
$\mathcal{H}_{{}_{|u\rangle}}$ we have the decomposition
\begin{equation}\label{decAS}
U|_{{}_{\mathcal{H}_{{}_{|u\rangle}}}} = 
\left\{ \begin{array}{lll}
\mathfrak{D}(\rho_0) \bigoplus \int \limits_{\rho>0} \mathfrak{S}(n=0, \rho) \, \ud \rho,
& \rho_0 = 1 - z_0,  z_0 = \frac{e^2}{\pi}, & \textrm{if} \,\, 0 < \frac{e^2}{\pi} <1  \\
\int \limits_{\rho>0} \mathfrak{S}(n=0, \rho) \, \ud \rho, & &
\textrm{if} \,\, 1 < \frac{e^2}{\pi}, 
\end{array} \right.
\end{equation}
into the direct integral of the unitary irreducible representations of the principal series
representations $\mathfrak{S}(n=0, \rho)$, with real $\rho>0$ and $n = 0$, and a discrete direct
summand of the supplementary series $\mathfrak{D}(\rho_0)$ corresponding to the value of the parameter
\[
\rho_0 = 1 - z_0,  z_0 = \frac{e^2}{\pi};
\]
and where $\ud \rho$ is the ordinary Lebesgue measure on $\mathbb{R}_+$.

Note that the irreducible unitary representations $\mathfrak{S}(n, \rho)$ of the principal series 
correspond to the representations $(l_0 = \frac{n}{2}, l_1 = \frac{i\rho}{2})$, with $n \in \mathbb{Z}$
and $\rho \in \mathbb{R}$ in the notation of the book \cite{Geland-Minlos-Shapiro}, and correspond to the character
$\chi = (n_1, n_2) = \big(\frac{n}{2} + \frac{i\rho}{2}, - \frac{n}{2} + \frac{i\rho}{2}\big)$ in the notation
of the book \cite{GelfandV}, and finally to the irreducible unitary representations
\[
U^{{}^{\chi_{{}_{n,\rho}}}} = \mathfrak{S}(n, \rho)
\]
induced by the unitary representations of the diagonal subgroup corresponding to the unitary character
$\chi_{{}_{n,\rho}}$ of the diagonal subgroup of $SL(2, \mathbb{C})$ within the Mackey theory of induced 
representations.

And recall that the irreducible unitary representations $\mathfrak{D}(\rho)$ of $SL(2, \mathbb{C})$ of the supplementary 
series are numbered by the real parameter $0<\rho <1$, and correspond to the representations
$(l_0 = 0, l_1 = \rho)$ in the notation of the book \cite{Geland-Minlos-Shapiro}. They also correspond to the character
$\chi = (n_1, n_2) = \big(\rho, \rho)$ in the notation
of the book \cite{GelfandV}, and finally to the irreducible unitary representations
\[
U^{{}^{\chi_{{}_{\rho}}}} = \mathfrak{D}(\rho)
\]
induced by the (non-unitary) representations of the diagonal subgroup of $SL(2, \mathbb{C})$ corresponding to 
the non-unitary character
$\chi_{{}_{\rho}}$ of the diagonal subgroup of $SL(2, \mathbb{C})$ within the Mackey theory of induced 
representations.

Next for each integer $m \in \mathbb{Z}$ and a point $u$ in the Lobachevsky space we consider spherically 
symmetric unit state vector $|m,u\rangle \in \mathcal{H}_{{}_{m}}$ 
\[
|m,u\rangle = e^{-imS(u)} |0\rangle
\] 
in the Hilbert space of the quantum field $S$.
If $u$ is the four-velocity of the reference frame in which the partial waves $f_{lm}^{(+)}$ 
are computed, then in this reference frame
\[
|m,u\rangle = e^{-imS_0}|0\rangle
\]
up to an irrelevant phase factor. 
The unit vector $|m,u\rangle$  has the following properties
\begin{enumerate}
\item[1m)]
$|m,u\rangle$ is an eigenstate of the total charge $Q$: $Q|u\rangle = em |m,u\rangle$.
\item[2m)]
$|m,u\rangle$ is spherically symmetric in the rest frame of $u$: 
$\epsilon^{\alpha \beta \mu \nu}u_{\beta} M_{\mu \nu} |m,u\rangle  = 0$, where $M_{\mu \nu}$
are the generators of the $SL(2, \mathbb{C})$ group.
\item[3m)]
$|m,u\rangle$ does not contain the (infrared) transversal photons: $N(u) |m,u\rangle = 0$.
\end{enumerate}
Proceeding exactly as Staruszkiewicz in \cite{Staruszkiewicz} (compare also  \cite{Staruszkiewicz2002})
we show that for any two points $u,v$ in the Lobachevsky space of unit time like four vectors 
\[
\langle u,m|m, v \rangle = \exp\Big\{ - \frac{e^2m^2}{\pi}(\lambda \textrm{coth} \lambda - 1) \Big\},
\]
where $\lambda$ is the hyperbolic angle between $u$ and $v$. Next, we construct the Hilbert subspace
$\mathcal{H}_{{}_{|m,u\rangle}} \subset \mathcal{H}_{{}_{m}}$ spanned by
\[
U_\alpha |m, u\rangle, \,\,\, \alpha \in SL(2, \mathbb{C}).
\]
Note that $\mathcal{H}_{{}_{|m,u\rangle}} \neq \mathcal{H}_{{}_{m}}$. Using the Gelfand-Neumark Fourier analysis
on the Lobachevsky space as Staruszkiewicz in \cite{Staruszkiewicz1992ERRATUM} we show that
\begin{equation}\label{decASm}
U|_{{}_{\mathcal{H}_{{}_{|m,u\rangle}}}} = 
\left\{ \begin{array}{lll}
\mathfrak{D}(\rho_0) \bigoplus \int \limits_{\rho>0} \mathfrak{S}(n=0, \rho) \, \ud \rho,
& \rho_0 = 1 - z_0,  z_0 = \frac{e^2 m^2}{\pi}, & \textrm{if} \,\, 0 < \frac{e^2 m^2}{\pi} <1  \\
\int \limits_{\rho>0} \mathfrak{S}(n=0, \rho) \, \ud \rho, & &

\textrm{if} \,\, 1 < \frac{e^2 m^2}{\pi}, 
\end{array} \right.
\end{equation}
where $\ud \rho$ is the Lebesgue measure on $\mathbb{R}_+$.

We need two Lemmata concerning the structure of the representation $U$ of $SL(2,\mathbb{C})$
in the Hilbert space of the quantum phase field $S$.

\vspace*{0.5cm}

LEMMA.
\[
U|_{{}_{\mathcal{H}_{{}_{m=1}}}} = U|_{{}_{\mathcal{H}_{{}_{|u\rangle}}}} \otimes U|_{{}_{\mathcal{H}_{{}_{m=0}}}}.
\]
 
\qedsymbol \,
First we show that (all tensor products in this Lemma are the Hilbert-space tensor products) 
\begin{equation}\label{H0timesHu=H1}
\mathcal{H}_{{}_{m=1}} = \mathcal{H}_{{}_{|u\rangle}} \otimes \mathcal{H}_{{}_{m=0}}
= \mathcal{H}_{{}_{|u\rangle}} \otimes \Gamma(\mathcal{H}_{{}_{m=0}}^{1})
\end{equation}
where $\mathcal{H}_{{}_{m=0}}^{1}$ is the single particle subspace of infrared transversal photons spanned by
\[
c_{lm}^{+} |0\rangle,
\]
and $\Gamma(\mathcal{H}_{{}_{m=0}}^{1})$ stands for the boson Fock space over $\mathcal{H}_{{}_{m=0}}^{1}$,
i.e. direct sum of symmetrized tensor products of $\mathcal{H}_{{}_{m=0}}^{1}$. 
The Hilbert subspace $\mathcal{H}_{{}_{|u\rangle}}$ is spanned by $|u\rangle$, and all its transforms
$U_{{}_{\Lambda(\alpha)}}|u\rangle = |u'\rangle$ with $u' = \Lambda(\alpha)^{-1}u$
ranging over the Lobachevsky space $\mathscr{L}_3 \cong SL(2,\,\mathbb{C})/SU(2,\mathbb{C})$ of 
time like unit four-vectors $u'$ -- the Lorentz images of the fixed $u$. 
The Hilbert space structure of $\mathcal{H}_{{}_{|u\rangle}}$ can be regarded as the one
induced by the invariant kernel
\[
u \times v \mapsto \langle u| v \rangle = \exp\Big\{ - \frac{e^2}{\pi}(\lambda \textrm{coth} \lambda - 1) \Big\},
\]
on the Lobachevsky space $\mathscr{L}_3$ as the RKHS corresponding to the kernel, compare e.g. \cite{PaulsenRaghupathi}. Because this kernel is continuous as a map $\mathscr{L}_3 \times \mathscr{L}_3
\mapsto \mathbb{R}$, and the Lobachevsky space is separable, then it is easily seen that there exists a denumerable
 subset $\{u_1, u_2, \ldots \} \subset \mathscr{L}_3$ such that $|u_1\rangle, u_2\rangle, \ldots$ are linearly independent and such that the denumerable set of finite 
rational (with $b_i \in \mathbb{Q}$) linear combinations
\[
\sum_{i=1}^{k} b_i |u_i\rangle
\]
of the elements $|u_1 \rangle, |u_2 \rangle, \ldots$ is dense in $\mathcal{H}_{{}_{|u\rangle}}$, compare e.g. 
\cite{Sikorski} Chap. XIII, \S 3. One can choose (Schmidt orthonormalization, \cite{Sikorski}, Chap XIII, \S 3)
out of them a denumerable and orthonormal system
\[
e_k(b_{1k}u_1, \ldots, b_{kk}u_k) = \sum_{i=1}^{k} b_{ik} |u_i\rangle 
= \sum_{i=1}^{k} b_{ik} e^{-iS(u_i)}|0\rangle, \,\,\, k=1,2, \ldots,
\]
wich is complete in $\mathcal{H}_{{}_{|u\rangle}}$. Note that
\[
U_{{}_{\Lambda(\alpha)}}|u\rangle = U_{{}_{\Lambda(\alpha)}}e^{-iS(u)} |0\rangle = U_{{}_{\Lambda(\alpha)}}e^{-iS(u)} 
U_{{}_{\Lambda(\alpha)}}^{-1} |0\rangle = e^{-iS(u')} |0\rangle 
\]
where $u' = \Lambda(\alpha)^{-1}u$ is the Lorentz image $u'$ in the Lobachevsky space of $u$ under the Lorentz
transformation $\Lambda(\alpha)$, because $|0\rangle$ is Lorentz invariant: $U|0\rangle = |0\rangle$.
In particular 
\begin{multline*}
U_{{}_{\Lambda(\alpha)}}e_k(b_{1k}u_1, \ldots, b_{kk}u_k) = e_k(b_{1k}u'_1, \ldots, b_{kk}u'_k), \\
= U_{{}_{\Lambda(\alpha)}} \big(\sum_{i=1}^{k} b_{ik} e^{-iS(u_i)}|0\rangle\big) 
= \sum_{i=1}^{k} b_{ik} e^{-iS(u'_i)}|0\rangle,
 \,\, u'_i = \Lambda(\alpha)^{-1}u_i, \,\,\, k=1,2,3, \ldots,
\end{multline*} 
forms another orthonormal and complete system in $\mathcal{H}_{{}_{|u\rangle}}$. In particular 
if $y \in \mathcal{H}_{{}_{|u\rangle}}$ then for some sequence of numbers $b^k \in \mathbb{C}$ such that
\[
||y||^2 = \sum_k |b^k|^2 < +\infty
\]
we have
\begin{equation}\label{yinHu}
y = \sum_{k = 1,2, \ldots} b^k e_k(b_{1k}u_1, \ldots, b_{kk}u_k) = 
 \sum_{k = 1,2, \ldots, i= 1, \ldots, k} b^k b_{ik} e^{-iS(u_i)}|0\rangle
\end{equation}
and
\[
U_{{}_{\Lambda(\alpha)}} y = \sum_{k = 1,2, \ldots} b^k e_k(b_{1k}u'_1, \ldots, b_{kk}u'_k) =
\sum_{k = 1,2, \ldots, i= 1, \ldots, k} b^k b_{ik} e^{-iS(u'_i)}|0\rangle.
\]

Similarly let us write shortly 
\[
c_{lm}^{+} = c_{\alpha}^{+} \,\,\, \textrm{and} \,\,\,
U_{{}_{\Lambda(\alpha)}} c_{lm}^{+} U_{{}_{\Lambda(\alpha)}}^{-1} = {c'}_{lm}^{+}.
\]
Then if $x \in \Gamma(\mathcal{H}_{{}_{m=0}}^{1}) = \mathcal{H}_{{}_{m=0}}$, then there exists
a multi-sequence of numbers $a^{\alpha_1 \ldots \alpha_n} \in \mathbb{C}$ such that
\[
|| x ||^2 = \sum_{n =1,2, \ldots, \alpha_1, \ldots, \alpha_n} (4\pi e^2)^n
\big| a^{\alpha_1 \ldots \alpha_n} \big|^2  < + \infty
\]
and 
\begin{equation}\label{xinH0}
x = \sum_{n =1,2, \ldots, \alpha_1, \ldots, \alpha_n}
a^{\alpha_1 \ldots \alpha_n} c_{\alpha_1}^{+} \ldots c_{\alpha_n}^{+} |0\rangle 
\end{equation}
\[
U_{{}_{\Lambda(\alpha)}} x = \sum_{n =1,2, \ldots, \alpha_1, \ldots, \alpha_n}
a^{\alpha_1 \ldots \alpha_n} {c'}_{\alpha_1}^{+} \ldots {c'}_{\alpha_n}^{+} |0\rangle \,\,\,
\]
where we have shortly written $\alpha_i$ for the pair $l_i, m_i$ with $-l_i \leq m_i \leq l_i$.

Before giving the definition of $x \otimes y$ for any general elements $x,y$ of the form 
(\ref{xinH0}) and respectively  (\ref{yinHu}) giving the
 algebraic tensor product  
$\mathcal{H}_{{}_{m=0}} \widehat{\otimes} \mathcal{H}_{{}_{|u\rangle}}$ densely included in
$\mathcal{H}_{{}_{m=1}}$, we need some further preliminaries. Namely note that 
the operators $c_{lm} = c_\alpha$ depend on the reference frame. For the construction of
$\otimes$ we need the operators in several reference frames. If the time-like axis of the referece frame has the unit versor $v \in \mathscr{L}_3$, then for the operator $c_\alpha = c_{lm}$ computed in this reference frame
we will write 

\[
~^v\!c_{\alpha} \,\,\, \textrm{or} \,\,\, ~^v\!c_{lm}
\] 
and 
\[
~^v\!c_{\alpha}^{+} \,\,\, \textrm{or} \,\,\, ~^v\!c_{lm}^{+}
\]
for their adjoints. Only for the fixed vector $u \in \mathscr{L}_3$ we simply write
\[
~^u\!c_{\alpha} = c_{\alpha}^{+} \,\,\, \textrm{or} \,\,\, ~^u\!c_{lm} = c_{lm}
\] 
and
\[
~^u\!c_{\alpha}^{+} = c_{\alpha}^{+} \,\,\, \textrm{or} \,\,\, ~^u\!c_{lm}^{+} = c_{lm}^{+}
\]
in order to simplify notation.

Now let 
\[
\overset{u \mapsto v}{A}_{\alpha \beta}
\]
be the unitary matrix transforming the orthonormal basis vectors $c_{\alpha}^{+} |0\rangle
= ~^u\!c_{\alpha}^{+} |0\rangle$ 
in $\mathcal{H}_{{}_{m=0}}$ 
\begin{equation}\label{transformation-cbeta+Vacuum}
~^v\!c_{\alpha}^{+} |0\rangle= 
\sum_{\beta} \overset{u \mapsto v}{A}_{\alpha \beta} ~^u\!c_{\beta}^{+}
|0\rangle =
\sum_{\beta} \overset{u \mapsto v}{A}_{\alpha \beta} c_{\beta}^{+}|0\rangle,
\end{equation}
under the Lorentz transformation $\Lambda_{uv}(\lambda_{uv})$ transforming the reference frame time-like versor
$u \in \mathscr{L}_3$ into the reference frame unit time-like versor $v \in \mathscr{L}_3$. In particular it 
gives the irreducible representation of the $SL(2, \mathbb{C})$ group in the single particle Hilbert subspace 
$\mathcal{H}_{{}_{m=0}}^{1}$ of infrared transversal photons spanned by
\[
c_{\alpha}^{+} |0\rangle = ~^u\!c_{\alpha}^{+} |0\rangle,
\]
and equal to the Gelfand-Minlos-Shapiro irreducible unitary representation 
$(l_0 = 1, l_1 = 0) = \mathfrak{S}(n=2, \rho= 0)$, computed explicitly in \cite{Staruszkiewicz1995}. 
Then, as shown in \cite{Staruszkiewicz1992}, it follows that
\begin{multline}\label{transformation-vcalpha}
U_{{}_{\Lambda_{uv}(\lambda_{uv})}} ~^u\!c_{\alpha} U_{{}_{\Lambda_{uv}(\lambda_{uv})}}^{-1}
= U_{{}_{\Lambda_{uv}(\lambda_{uv})}} c_{\alpha} U_{{}_{\Lambda_{uv}(\lambda_{uv})}}^{-1}
= ~^v\!c_{\alpha} = \\
\sum_{\beta} \overline{\overset{u \mapsto v}{A}_{\alpha \beta}} ~^u\!c_{\beta} 
+ \overline{\overset{u \mapsto v}{B}_\alpha} \, Q  \\ =
\sum_{\beta} \overline{\overset{u \mapsto v}{A}_{\alpha \beta}} c_{\beta} 
+ \overline{\overset{u \mapsto v}{B}_\alpha} \, Q,
\end{multline}
and\footnote{We are using slightly different convention than \cite{Staruszkiewicz1992}, with ours 
$\overset{u \mapsto v}{A}_{\alpha \beta}$ corresponding to the complex conjugation $\overline{A_{\alpha \beta}}$ of the matrix elements 
$A_{\alpha \beta}$ used in  \cite{Staruszkiewicz1992} and similarly our numbers 
$\overset{u \mapsto v}{B}_\alpha$ correspond to the complex conjugation $\overline{B_\alpha}$ of the numbers 
$B_\alpha$ used in \cite{Staruszkiewicz1992}.}
\begin{multline}\label{transformation-S}
U_{{}_{\Lambda_{uv}(\lambda_{uv})}} S(u) U_{{}_{\Lambda_{uv}(\lambda_{uv})}}^{-1}
= S(v) = \\
S(u) + \frac{1}{4\pi i e}\sum_{\alpha \beta} \big( \overset{u \mapsto v}{B}_\alpha\overline{\overset{u \mapsto v}{A}_{\alpha \beta}} ~^u\!c_{\beta} -
\overline{\overset{u \mapsto v}{B}_\alpha} \overset{u \mapsto v}{A}_{\alpha \beta} ~^u\!c_{\beta}^{+} \big) 
\end{multline}
and thus
\begin{multline}\label{transformation-vcalpha+}
U_{{}_{\Lambda_{uv}(\lambda_{uv})}} ~^u\!c_{\alpha}^{+} U_{{}_{\Lambda_{uv}(\lambda_{uv})}}^{-1}
= U_{{}_{\Lambda_{uv}(\lambda_{uv})}} c_{\alpha}^{+} U_{{}_{\Lambda_{uv}(\lambda_{uv})}}^{-1}
= ~^v\!c_{\alpha}^{+} = \\
\sum_{\beta} \overset{u \mapsto v}{A}_{\alpha \beta} ~^u\!c_{\beta}^{+} 
+ \overset{u \mapsto v}{B}_\alpha Q \\ =
\sum_{\beta} \overset{u \mapsto v}{A}_{\alpha \beta} c_{\beta}^{+} 
+ \overset{u \mapsto v}{B}_\alpha Q,
\end{multline}
where $Q$ is the charge operator and where $\overset{u \mapsto v}{B}_\alpha$ are complex numbers depending
on the transformation $\Lambda_{uv}(\lambda_{uv})$ mapping $u \mapsto v = \Lambda_{uv}(\lambda_{uv})^{-1}u$ such that 
\[
\sum_{\alpha} |\overset{u \mapsto v}{B}_\alpha|^2 = 8e^2 (\lambda_{uv} \textrm{coth} \lambda_{uv} -1)
\]
with $\lambda_{uv}$ equal to the hyperbolic angle between $u$ and $v$. Note that the charge operator 
is invariant (commutes with $U_{{}_{\Lambda_{uv}(\lambda_{uv})}}$) and is identical in each 
reference frame so that no superscript $u$ nor $v$ is needed for $Q$.

The limit on the right hand side of the equality (\ref{transformation-cbeta+Vacuum}) should be understood 
in the sense of the ordinary Hilbert space norm
in the Hilbert space of the quantum phase field $S$. In general all limits in the expressions containing linear combinations of operators acting on $|0\rangle$ should be understood in this manner.  

Now let us explain why for each fixed $\alpha$ we need essentially all $~^v\!c_{\alpha}$, $v \in \mathscr{L}_3$
for the construction of the bilinear map $x\times y \mapsto x\otimes y$ which serves to define the algebraic tensor product
$\mathcal{H}_{{}_{m=0}} \widehat{\otimes} \mathcal{H}_{{}_{|u\rangle}}$ of the Hilbert spaces $\mathcal{H}_{{}_{m=0}}$ 
and $\mathcal{H}_{{}_{|u\rangle}}$.
In particular consider two vectors $c_\alpha^{+} |0\rangle$ and $e^{-iS(v)}|0\rangle$ with $v$
not equal to the fixed time like versor $u$ of the reference frame in which the partial waves $f_{lm}^{(+)}$ 
and the operators $c_{lm} = c_{\alpha} = ~^uc_{\alpha}$ are computed. Perhaps it would be tempting to put
\[
c_\alpha^{+}e^{-iS(v)} |0\rangle
\]    
for the tensor product of $c_\alpha^{+} |0\rangle$ and $e^{-iS(v)}|0\rangle$, but this would be a wrong definition. In particular 
\begin{multline*}
\langle0|e^{iS(v)}~^u\!c_{\beta} ~^uc_{\alpha}^{+}e^{-iS(v)}|0\rangle
=
\langle0|e^{iS(v)}c_{\beta} c_{\alpha}^{+}e^{-iS(v)}|0\rangle \neq \\
 \neq 
\langle 0|~^uc_{\beta} ~^uc_{\alpha}^{+} |0\rangle
\langle0| e^{iS(v)}e^{-iS(v)}|0\rangle
=
\langle 0|c_{\beta} c_{\alpha}^{+} |0\rangle
\langle 0| e^{iS(v)}e^{-iS(v)}|0\rangle
\end{multline*}
contrary to what is expected of the inner product for simple tensors. 
This is mainly because $c_{\alpha} = ~^u\!c_{\alpha}$
do not commute with $e^{-iS(v)}$ for $u \neq v$. 
However for any two $u,w \in \mathscr{L}_3$, 
\begin{equation}\label{inn-prod-vacexp(S(v))c(v)c(w)+exp(S(w))vac}
\langle 0 |e^{iS(v)}~^v\!c_{\beta} ~^w\!c_{\alpha}^{+} e^{-iS(w)}|0\rangle
= \langle 0| ~^v\!c_{\beta} ~^w\!c_{\alpha}^{+} |0\rangle 
\langle 0| e^{iS(v)} e^{-iS(w)}|0 \rangle
\end{equation}  
which easily follows from (\ref{transformation-vcalpha})
- (\ref{transformation-vcalpha+}) and from the canonical commutation relations.
Similarly for the case when two (or more) creation operators are involved
\begin{multline}\label{inn-prod-vacexp(S(v))c(v)c(v)c(w)+c(w+)exp(S(w))vac}
\langle 0 |e^{iS(v)}~^v\!c_{\beta_1}~^v\!c_{\beta_2} ~^w\!c_{\alpha_1}^{+}  ~^w\!c_{\alpha_1}^{+} e^{-iS(w)}|0\rangle
= \langle 0| ~^v\!c_{\beta_1}~^v\!c_{\beta_2} ~^w\!c_{\alpha_1}^{+}  ~^w\!c_{\alpha_2}^{+}|0\rangle 
\langle 0| e^{iS(v)} e^{-iS(w)}|0 \rangle, \\
\langle 0 |e^{iS(v)}~^v\!c_{\beta_1} \ldots ~^v\!c_{\beta_n} 
~^w\!c_{\alpha_1}^{+} \ldots ~^w\!c_{\alpha_n}^{+} e^{-iS(w)}|0\rangle \\
= \langle 0| ~^v\!c_{\beta_1} \ldots ~^v\!c_{\beta_n} 
~^w\!c_{\alpha_1}^{+} \ldots ~^w\!c_{\alpha_n}^{+}|0\rangle 
\langle 0| e^{iS(v)} e^{-iS(w)}|0 \rangle
\end{multline}  
as expected of the inner product on simple tensors. This explains the need for using 
$~^v\!c_{lm} = ~^v\!c_{\alpha}$ in various reference frames $v$, as in composing any complete orthomnormal
system in $\mathcal{H}_{{}_{|u\rangle}}$ we need linear combinations of vectors 
\[
e^{-iS(v)}|0\rangle
\]
with various $v \in \mathscr{L}_3$.

Therefore for any $v \in \mathscr{L}_3$ we put
\begin{multline}\label{def-otimes-particular}
\big( ~^v\!c_{\alpha_1}^{+}  ~^v\!c_{\alpha_2}^{+}|0\rangle) \otimes
\big(  e^{-iS(v)}|0\rangle \big) =
 ~^v\!c_{\alpha_1}^{+}  ~^v\!c_{\alpha_2}^{+}e^{-iS(v)}|0\rangle, \\
\big( ~^v\!c_{\alpha_1}^{+} \ldots ~^v\!c_{\alpha_n}^{+}|0\rangle) \otimes
\big(  e^{-iS(v)}|0\rangle \big) =
 ~^v\!c_{\alpha_1}^{+} \ldots  ~^v\!c_{\alpha_n}^{+}e^{-iS(v)}|0\rangle.
\end{multline}

Let in particular $U$ be the unitary representor of a Lorentz transformation which transforms $v$ into $v'$.
Then 
\[
~^v\!c_{\alpha}^{+} = \\
\sum_{\beta} \overset{w \mapsto v}{A}_{\alpha \beta} ~^w\!c_{\alpha}^{+} 
+ \overset{w \mapsto v}{B}_\alpha Q
\]
and 
\begin{multline*}
(U ~^v\!c_{\alpha}^{+} |0\rangle) \otimes (U e^{-iS(w)}|0\rangle) =
(~^{v'}\!c_{\alpha}^{+} |0\rangle) \otimes (e^{-iS(w')}|0\rangle)  \\ =
\big( \sum_{\beta} \overset{w' \mapsto v'}{A}_{\alpha \beta} ~^{w'}\!c_{\alpha}^{+} |0\rangle \big) \otimes 
\big( e^{-iS(w')}|0\rangle \big) \\ =
\sum_{\beta} \overset{w' \mapsto v'}{A}_{\alpha \beta} ~^{w'}\!c_{\alpha}^{+} e^{-iS(w')}|0\rangle \\
= \sum_{\beta} \overset{w \mapsto v}{A}_{\alpha \beta} ~^{w'}\!c_{\alpha}^{+} e^{-iS(w')}|0\rangle \\
= U \big( \sum_{\beta} \overset{w \mapsto v}{A}_{\alpha \beta} ~^w\!c_{\alpha}^{+} e^{-iS(w)}|0\rangle \big),
\end{multline*}
so that
\[
(U ~^v\!c_{\alpha}^{+} |0\rangle) \otimes (U e^{-iS(w)}|0\rangle) =
U \big((~^v\!c_{\alpha}^{+} |0\rangle) \otimes (e^{-iS(w)}|0\rangle) \big)
\]
and similarly we show that this is the case for more general simple tensors 
\begin{equation}\label{U=UxU-on-simple-tensors}
\big( U \, ~^v\!c_{\alpha_1}^{+} \ldots ~^v\!c_{\alpha_n}^{+}|0\rangle) \otimes
\big(  U \, e^{-iS(v)}|0\rangle \big) = \\
U \Big( \big( ~^v\!c_{\alpha_1}^{+} \ldots ~^v\!c_{\alpha_n}^{+}|0\rangle) \otimes
\big(  e^{-iS(v)}|0\rangle \big) \Big).
\end{equation}

Now in order to define $x \otimes y$ for general $x,y$ of the form (\ref{xinH0}) and respectively  (\ref{yinHu})
we need to extend the formula (\ref{def-otimes-particular}). In fact $x \otimes y$ is uniquelly determined
by (\ref{def-otimes-particular}). Now we prepare the explicit formula for $x \otimes y$ out of 
(\ref{def-otimes-particular}).

Let $u_1, u_2, \ldots \in \mathscr{L}_3$ be the unit fourvectors which are used in the definition of the 
complete orthonormal system  
\[
e_k(b_{1k}u_1, \ldots, b_{kk}u_k) = \sum_{i=1}^{k} b_{ik} |u_i\rangle 
= \sum_{i=1}^{k} b_{ik} e^{-iS(u_i)}|0\rangle, \,\,\, k=1,2, \ldots,
\]
in $\mathcal{H}_{{}_{|u\rangle}}$. Corresponding to them we define 

\[
~^{u_i}\!c_{\alpha} = 
\sum_{\beta} \overline{\overset{u \mapsto u_i}{A}_{\alpha \beta}} \, ~^u\!c_{\alpha}
+ \overline{\overset{u \mapsto v}{B}_\alpha} \, Q  \\ =
\sum_{\beta} \overline{\overset{u \mapsto u_i}{A}_{\alpha \beta}} \, c_{\alpha} 
+ \overline{\overset{u \mapsto u_i}{B}_\alpha} \, Q,
\]
and 
\[
~^{u_i}\!c_{\alpha}^{+} = 
\sum_{\beta} \overset{u \mapsto u_i}{A}_{\alpha \beta} ~^u\!c_{\alpha}^{+} 
+ \overset{u \mapsto v}{B}_\alpha \, Q  \\ =
\sum_{\beta} \overset{u \mapsto u_i}{A}_{\alpha \beta} c_{\alpha}^{+} 
+ \overset{u \mapsto u_i}{B}_\alpha \, Q.
\]
Having defined this we introduce for each $i=1,2, \ldots$ and the corresponding
operator $~^{u_i}\!c_{\alpha}$ the operator
\begin{equation}\label{icalpha}
~^i\!c_{\alpha} = 
\sum_{\beta} \overline{\overset{u_i\mapsto u}{A}_{\alpha \beta}} ~^{u_i}\!c_{\alpha} 
\end{equation}
by discarding the part proportional to the total charge $Q$ in the operator
\[
c_{\alpha} = ~^u\!c_{\alpha} = 
\sum_{\beta} \overline{\overset{u_i\mapsto u}{A}_{\alpha \beta}} \, ~^{u_i}\!c_{\beta} 
+ \overline{\overset{u_i \mapsto u}{B}_\alpha} \, Q
\]
as obtained by the transformation $u_i \mapsto u$ transforming the system of operators $~^{u_i}\!c_{\beta}$
into the system of operators $~^{u}\!c_{\alpha}$.
Of course we have
\[
c_{\alpha}^{+}= ~^u\!c_{\alpha}^{+} = 
\sum_{\beta} \overset{u_i\mapsto u}{A}_{\alpha \beta} ~^{u_i}\!c_{\beta}^{+} 
+ \overset{u_i \mapsto u}{B}_\alpha \, Q.
\]

The crucial facts for the computations which are to follow are the following. 
For each four-vector $v \in \mathscr{L}_3$ 
\[
[~^v\!c_{\alpha}, e^{-iS(v)}] =0.
\]
The commutation rules are preserved and 
\[
[~^v\!c_{\alpha}, ~^v\!c_{\beta}] = 0, \,\,\,
[~^v\!c_{\alpha}, ~^v\!c_{\beta}^{+}] = 4\pi e^2 \, \delta_{{}_{\alpha \beta}}, \,\,\,
[Q, ~^v\!c_{\alpha}] = 0, \,\,\, ~^v\!c_{\alpha}|0\rangle = \langle0| ~^v\!c_{\alpha}^{+} = 0.
\]
But moreover, if we fix arbitrarily $\alpha = (l,m)$ then because the operators 
$~^i\!c_{\alpha}$, $i= 1,2,\ldots$ all differ from the fixed operator 
$c_{\alpha} = ~^u\!c_{\alpha}$ with fixed $u \in \mathscr{L}_3$ by the operator (depending on 
$i$) which is always proportional to the total charge operator $Q$, as a consequence of the transformation rule
(\ref{transformation-vcalpha}) and (\ref{transformation-vcalpha+}), then not only 
\[
[~^i\!c_{\alpha}, ~^i\!c_{\beta}] = 0, \,\,\,
[~^i\!c_{\alpha}, ~^i\!c_{\beta}^{+}] = 4\pi e^2 \, \delta_{{}_{\alpha \beta}}, \,\,\,
[Q, ~^i\!c_{\alpha}] = 0, \,\,\, ~^i\!c_{\alpha}|0\rangle = \langle0| ~^i\!c_{\alpha}^{+} = 0,
\,\,\, i= 1,2, \ldots
\]
for all $i = 1,2, \ldots$ but likewise 
\[
[~^i\!c_{\alpha}, ~^j\!c_{\beta}] = 0, \,\,\,
[~^i\!c_{\alpha}, ~^j\!c_{\beta}^{+}] = 4\pi e^2 \, \delta_{{}_{\alpha \beta}}, \,\,\,
[Q, ~^i\!c_{\alpha}] = 0, \,\,\, ~^i\!c_{\alpha}|0\rangle = \langle0| ~^i\!c_{\alpha}^{+} = 0,
\,\,\, i,j = 1,2, \ldots.
\]
Note also that 
\[
c_{\alpha}^{+}|0\rangle = ~^i\!c_{\alpha}^{+} |0\rangle, \,\,\, i = 1,2, 3, \ldots.
\]
Furthermore we have the following orthogonality relations
\begin{multline}\label{orthonormality-vacOps+ccccc+c+c+c+Opkvac}
\Big\langle 0 \Big| \Big(\sum_{j=1}^{s} b_{js} e^{iS(u_j)} ~^j\!c_{\beta_1} \ldots ~^j\!c_{\beta_{m}} \Big)
\Big(\sum_{i=1}^{k} b_{ik} ~^i\!c_{\alpha_1}^{+} \ldots  ~^i\!c_{\alpha_n}^{+} e^{-iS(u_i)} \Big) \Big| 0 \Big\rangle \\
= 
(4\pi e^2)^n \,
\delta_{sk} \, \delta_{mn} \, \delta_{{}_{\{\alpha_1 \ldots \alpha_n\} \,\,\,\{\beta_1 \ldots \beta_m\}}}.
\end{multline} 

Let $x,y$ be general elements respectively $x \in \mathcal{H}_{{}_{m=0}}$ and 
$y \in \mathcal{H}_{{}_{|u\rangle}}$ of the general form (\ref{xinH0}) and respectively  (\ref{yinHu}). 
We define the following bilinear map $\otimes$ of $\mathcal{H}_{{}_{m=0}} \times \mathcal{H}_{{}_{|u\rangle}}$ 
into $\mathcal{H}_{{}_{m=1}}$ by the formula
\begin{multline*}
x \times y \mapsto x \otimes y  \\ = 
\sum_{n =1,2, \ldots, k = 1,2, \ldots, i = 1, \ldots, k, \alpha_1, \ldots, \alpha_n}
a^{\alpha_1 \ldots \alpha_n} b^k b_{ik} ~^i\!c_{\alpha_1}^{+} \ldots ~^i\!c_{\alpha_n}^{+} e^{-iS(u_i)}|0\rangle.
\end{multline*}
We show now that $\mathcal{H}_{{}_{m=0}}$ and $\mathcal{H}_{{}_{|u\rangle}}$ are $\otimes$-linearly
disjoint \cite{treves}, compare Part III, Chap. 39, Definition 39.1. Namely let $y_1, \ldots, y_r$
be a finite subset of generic elements
\[
y_j = \sum_{k = 1,2, \ldots} b^{k}_{j} e_k(b_{1k}u_1, \ldots, b_{kk}u_k) = 
\sum_{k = 1,2, \ldots, i= 1, \ldots, k} b^{k}_{j} b_{ik} e^{-iS(u_i)}|0\rangle
\]
in $\mathcal{H}_{{}_{|u\rangle}}$ for $j = 1, \ldots, r$; and similarly let $x_1, \ldots, x_r$
be a finite subset of generic elements
\[
x_j = \sum_{n =1,2, \ldots, \alpha_1, \ldots \alpha_n}
a^{\alpha_1 \ldots \alpha_n}_{j} c_{\alpha_1}^{+} \ldots c_{\alpha_n}^{+} |0\rangle
\]
in $\mathcal{H}_{{}_{m=0}}$ for $j = 1, \ldots, r$. Let us suppose that 
\begin{multline}\label{x-tensor-y=0}
\sum_{j=1}^{r} x_j \otimes y_j \\ = 
\sum_{j=1,\ldots,r, n =1,2, \ldots, k = 1,2, \ldots, i = 1, \ldots, k, \alpha_1, \ldots, \alpha_n}
a^{\alpha_1 \ldots \alpha_n}_{j} b^{k}_{j} b_{ik} ~^i\!c_{\alpha_1}^{+} \ldots ~^i\!c_{\alpha_n}^{+} e^{-iS(u_i)}|0\rangle = 0,
\end{multline}
and that $x_1, \ldots, x_r$ are linearly independent. We have to show that $y_1 = \ldots = y_r = 0$.
The linear inependence of $x_j$ means that if for numbers $b^j$ it follows that
\[
\sum_{j=1}^{r} b^j a^{\alpha_1 \ldots \alpha_n}_{j} = 0
\] 
for all $n= 1,2, \ldots$, $\alpha_i = (1,-1), (1,0), (1,1), (2,-2), \ldots$
then $b_1 = \ldots = b_r = 0$.
Now consider the inner product of the left hand side
of (\ref{x-tensor-y=0}) with 
\[
\sum_{q=1}^{k} b_{qk} ~^q\!c_{\beta_1}^{+} \ldots ~^q\!c_{\beta_n}^{+} e^{-iS(u_q)}|0\rangle. 
\]
Then from (\ref{x-tensor-y=0}) and the orthogonality relations (\ref{orthonormality-vacOps+ccccc+c+c+c+Opkvac}) 
we get
\[
\sum_{j = 1}^{r}
a^{\beta_1 \ldots \beta_n}_{j} b^{k}_{j} = 0
\]
for each $k = 1,2, \ldots$. Therefore by the linear independence of $x_j$ we obtain
\[
b^{k}_{1} = \ldots = b^{k}_{r} = 0
\]
for each $k = 1,2, \ldots$, so that 
\[
y_1 = \ldots = y_r = 0.
\]
Similarly from (\ref{x-tensor-y=0}) and linear independence of $y_1, \ldots, y_r$ it follows
that 
\[
x_1 = \ldots = x_r = 0,
\]
so that  $\mathcal{H}_{{}_{m=0}}$ and $\mathcal{H}_{{}_{|u\rangle}}$ are $\otimes$-linearly
disjoint. 

By construction the image of  $ \otimes: \mathcal{H}_{{}_{m=0}} \times \mathcal{H}_{{}_{|u\rangle}}
\rightarrow \mathcal{H}_{{}_{m=1}}$ span the Hilbert space $\mathcal{H}_{{}_{m=1}}$ and is dense
in $\mathcal{H}_{{}_{m=1}}$. 
Therefore the image of $\otimes$ defines the algebraic tensor product
$\mathcal{H}_{{}_{m=0}} \otimes_{{}_{\textrm{alg}}} \mathcal{H}_{{}_{|u\rangle}}$ of 
$\mathcal{H}_{{}_{m=0}}$ and $\mathcal{H}_{{}_{|u\rangle}}$ densely included in 
$\mathcal{H}_{{}_{m=1}}$. 

Now we show that the inner product $\langle \cdot | \cdot \rangle$ on $\mathcal{H}_{{}_{m=1}}$, if restricted to the algebraic tensor product subspace
$\mathcal{H}_{{}_{m=0}} \otimes_{{}_{\textrm{alg}}} \mathcal{H}_{{}_{|u\rangle}}$, coincides with the 
inner product of the algebraic Hilbert space tensor product: 
\[
\langle x\otimes y |x' \otimes  y'\rangle = \langle x|x'\rangle \langle y | y' \rangle
\]
for any  generic elements $x,x' \in \mathcal{H}_{{}_{m=0}}$ and any generic elements 
$y,y' \in \mathcal{H}_{{}_{|u\rangle}}$. Indeed let $x,y$ be generic elements of the form
(\ref{xinH0}) and (\ref{yinHu}) respectively and similarly for the generic elements
$x',y'$ we put
\[
x' = \sum_{q =1,2, \ldots, \beta_1, \ldots, \beta_q}
a'^{\beta_1 \ldots \beta_n} c_{\beta_1}^{+} \ldots c_{\beta_q}^{+} |0\rangle 
\]
and 
\[
y' = \sum_{s = 1,2, \ldots} b'^{s} e_s(b_{1s}u_1, \ldots, b_{ss}u_s)
= \sum_{s = 1,2, \ldots, j= 1, \ldots, s} b'^{s} b_{js} e^{-iS(u_j)}|0\rangle.
\]
Then 
\begin{multline*}
\langle x' \otimes y' |x \otimes  y\rangle = \sum_{n,k, q, s, \alpha_1, \ldots, \alpha_n, \beta_1, \ldots, \beta_q} 
\overline{a'^{\beta_1 \ldots \beta_q}}a^{\alpha_1 \ldots \alpha_n}  \overline{b'^{s}}b^k  \,\,\times
\\ \times \,\,
\Big\langle 0 \Big| 
 \Big(\sum_{j=1}^s b_{js} e^{iS(u_j)} ~^j\!c_{\beta_q} \ldots ~^j\!c_{\beta_1} \Big) 
\Big(\sum_{i=1}^{k}~^i\!c_{\alpha_1}^{+} \ldots ~^i\!c_{\alpha_n}^{+}e^{-iS(u_i)} \Big)
\Big|0 \Big\rangle
\end{multline*}
which, on using (\ref{inn-prod-vacexp(S(v))c(v)c(v)c(w)+c(w+)exp(S(w))vac}) and the orthogonality
relations (\ref{orthonormality-vacOps+ccccc+c+c+c+Opkvac}),
is equal to 
\[
\Big(\sum_{n,\alpha_1, \ldots \alpha_n} (4 \pi e^2)^n \,
\overline{a'^{\alpha_1 \ldots \alpha_n}}a^{\alpha_1 \ldots \alpha_n} \Big) 
\Big( \sum_{k}
 \overline{b'^{k}}b^k \Big) =  \langle x|x'\rangle \langle y | y' \rangle.
\]
Thus the proof of the equality (\ref{H0timesHu=H1}) is now complete. 

Now let $x,y$ be any generic elements of the form (\ref{xinH0}) and (\ref{yinHu}) respectively.
Then by repeated application of (\ref{U=UxU-on-simple-tensors}) and the continuity of each representor\footnote{Each representor $U_{{}_{\Lambda(\alpha)}}$ being unitary is bounded and thus continuous in the 
topology of the Hilbert space.} $U$ we obtain 
\[
U(x \otimes y) = Ux \otimes Uy.
\]
This ends the proof of our Lemma.
\qed

We observe now that the same proof can be repeated in showing validity of the following

\vspace*{0.5cm}

LEMMA.
\[
U|_{{}_{\mathcal{H}_{{}_{m}}}} = U|_{{}_{\mathcal{H}_{{}_{|m,u\rangle}}}} \otimes U|_{{}_{\mathcal{H}_{{}_{m=0}}}}.
\]
 
\qed

Now let\footnote{Note that the standard 
definition of the integer part is slightly different.}  `$\textrm{Integer part} \, x$' for any positive real number $x$ be the least natural number among
all natural numbers $n$ for which $x \leq n$.
Joining the last Lemma with the result (\ref{decASm}) of Staruszkiewicz \cite{Staruszkiewicz1992ERRATUM}
we obtain the theorem formulated in Introduction.

\vspace*{0.5cm}

The author is indebted for helpful discussions to prof. A. Staruszkiewicz.




\begin{thebibliography}{99}


\vspace{.5cm}


\bibitem{Geland-Minlos-Shapiro} Gelfand, I. M., Minlos, R. A,, Shapiro, Z. Ya.: Representations of the rotation 
and Lorentz groups and their applications. Pergamon Press Book, The Macmillan Company, New York, 1963.
\bibitem{GelfandV} Gelfand, I. M., Graev, M. I. and Vilenkin, N. Ya.: Generalized Functions. Vol V. Academic Press, New York and London, 1966.
\bibitem{PaulsenRaghupathi} Paulsen, V. I., Raghupathi, M.: An introduction to the theory of reproducing kernel Hilbert
spaces, Cambridge University Press, Cambridge 2016.
\bibitem{Sikorski} Sikorski, R.: Real functions, Vol II, PWN, Warszawa (1959) (in Polish).
\bibitem{Staruszkiewicz1987} Staruszkiewicz, A.: Quntum mechanics of phase and charge and quantization
of the Coulomb field. Preprint TPJU-12/87, June 1987.
\bibitem{Staruszkiewicz} Staruszkiewicz, A.: Ann. Phys. (N.Y.) 190, 354 (1989).
\bibitem{Staruszkiewicz1992} Staruszkiewicz, A.: Acta Phys. Polon. {\bf B 23}, 927 (1992).
\bibitem{Staruszkiewicz1992ERRATUM} Staruszkiewicz, A.: Acta Phys. Polon. {\bf B23}, 591 (1992)
and ERRATUM in Acta Phys. Pol {\bf B23}, 959 (1992).
\bibitem{Staruszkiewicz1995} Staruszkiewicz, A.: Acta Phys. Polon. {\bf B26}, 1275 (1995).
\bibitem{Staruszkiewicz2002} Staruszkiewicz, A.: Foundations of Physics {\bf 32}, 1863 (2002).
\bibitem{Staruszkiewicz2004} Staruszkiewicz, A.: Acta Phys. Polon. {\bf B35}, 2249 (2004).
\bibitem{Staruszkiewicz2009} Staruszkiewicz, A.: Reports on Math. Phys. {\bf 64}, 293 (2009).
\bibitem{treves} Treves, F.: Topological vector spaces, distributions and kernels. Academic Press, 1967.




\end{thebibliography}
\end{document}